\documentclass[journal]{IEEEtran}
\usepackage{cite,graphicx,amsmath,amssymb,bm}
\usepackage[usenames,dvipsnames]{xcolor}

\newcommand{\figref}{Fig.~\ref}

\newcommand{\ve}[1]{\mathbf{#1}}

\newcommand{\uve}[1]{\mathbf{\hat{{#1}}}}
\newcommand{\uvs}[1]{\boldsymbol{\hat{#1}}}

\ifCLASSINFOpdf
\else
\fi

\begin{document}

\title{Unidirectional Loop Metamaterials (ULM) \\
as Magnetless Artificial Ferrimagnetic Materials: \\ Principles and Applications\vspace{-2mm}
}
\author{Toshiro Kodera,~\IEEEmembership{Senior Member,~IEEE,} and
        Christophe~Caloz,~\IEEEmembership{Fellow,~IEEE}
\thanks{T. Kodera is with the Department
of Electrical Engineering, Meisei University,
Tokyo Japan (e-mail: toshiro.kodera@meisei-u.ac.jp).
C. Caloz is with the Department
of Electrical and Engineering, \'{E}cole Polytechnique de
Montr\'{e}al, Montr\'{e}al,
QC, H2T 1J3 Canada.}
\thanks{}
}
\markboth{}
{Shell \MakeLowercase{\textit{et al.}}: Bare Demo of IEEEtran.cls for IEEE Journals}
\maketitle

\begin{abstract}
This paper presents an overview of Unidirectional Loop Metamaterial (ULM) structures and applications. Mimicking electron spin precession in ferrites using loops with unidirectional loads (typically transistors), the ULM exhibits all the fundamental properties of ferrite materials, and represents the \emph{only existing magnetless ferrimagnetic medium}. We present here an extended explanation of ULM physics and unified description of its component and system applications.
\end{abstract}
\vspace{-2mm}


\vspace{-4mm}
\section{Introduction}
\vspace{-1mm}

Over the past decades, nonreciprocal components (isolators, circulators, nonreciprocal phase shifters, etc.) have been have been almost exclusively implemented in ferrite technology~\cite{ferrite_BSTJ_1955, Matthaei_gyrator_1959, Lax_1962, Optical_Faraday_iso_1964, Optical_iso_Wang_1971, Rodrigue_1988, YIG_faraday_rotator_1983, Pozar_ME_2011}. This has been the case in both microwaves and optics, despite distinct underlying physics, namely the purely magnetic effect (electron spin precession) in the former case~\cite{Faraday_1933,Lax_1962,Gurevich_1996} and the magneto-optic effect (electron cyclotron orbiting)~\cite{Rayleigh_1901,Landau_1984,Zvezdin_Kotov_1997} in the latter case. However, ferrite components suffer from the well-known issues high-cost, high-weight and incompatibility with integrated circuit technology, and \emph{magnetless nonreciprocity} has therefore long been consider a holy grail in this area~\cite{Caloz_AWPL_NR_I_2018, Caloz_AWPL_NR_II_2018}.

There have been several attempts to develop magnetless nonreciprocal components, specifically 1)~\emph{active circuits}~\cite{Tanaka_circulator_1965, Ayasli_circulator_1989, Bahl_6port_circulator_1988, YH_Wang_auasi_circ_2016, Kiaei_quasi_circ_2017, Fang_quasi_circ_2017}, and space-time~\cite{Caloz_AWPL_NR_II_2018} 2)~\emph{modulated structures}~\cite{Ethan_Wang_modu_cap_2014, Alu_circulator_2016, Alu_2018, Taravati_PRB_10_2017, Taravati_2016_2,Alu_2016} and 3)~\emph{switched structures}~\cite{Harish_Staggered_com_2016} (both based on 1950ies parametric (e.g. \cite{Cullen_1958,Landauer_1960}) or commutated (e.g.~\cite{Franks_SW_CAP_1960}) microwave systems). All have their specific features, as indicated in Tab.~\ref{tab:isol_FET_cp}.

\vspace{-4mm}
\begin{table}[ht]
\centering
\caption{Comparison (typical and relative terms) between different magnetless nonreciprocity technologies plus ferrite.}
\vspace{-2mm}
\begin{tabular}{p{13mm}p{9mm}p{9mm}p{8mm}p{9mm}p{8mm}}
&material&{$P_\text{consum.}$}&{bias}&{cost}&{noise}\\
\hline\hline
ferrite&yes&zero&magnet&high&N/A\\
\hline
act. circ.&no&low&DC&low&low\\
switched&no&med.&RF&high&high\\
modulated&no&med.&RF&med.&med.\\
\textbf{{ULM}}&\textbf{YES}&\textbf{low}&\textbf{DC}&\textbf{low}&\textbf{med}\\
\end{tabular}
\label{tab:isol_FET_cp}
\end{table}
\vspace{-3mm}

We introduced in 2011~\cite{Kodera_APL_2011} in a Unidirectional Loop Metamaterial (ULM) mimicking ferrites at microwaves and representing the only artificial ferrite material, or metamaterial, existing to date. This paper presents an overview of the ULM and its applications reported to date.

\section{Operation Principle}
\vspace{-1mm}

A Unidirectional Loop Metamaterial (ULM) may be seen as a \emph{physicomimetic\footnote{The adjective ``physicomimetic'' is meant here, from etymology, as ``mimicking physics.''} artificial implementation of a ferrite in the microwave regime}. Its operation principle is thus based on \emph{microscopic unidirectionality}, from which the \emph{macroscopic description} is inferred upon averaging.

\subsection{Microscopic Description}\label{sec:micro_descr}

\begin{figure}[htp]
\centering
\includegraphics[width=0.4\textwidth]{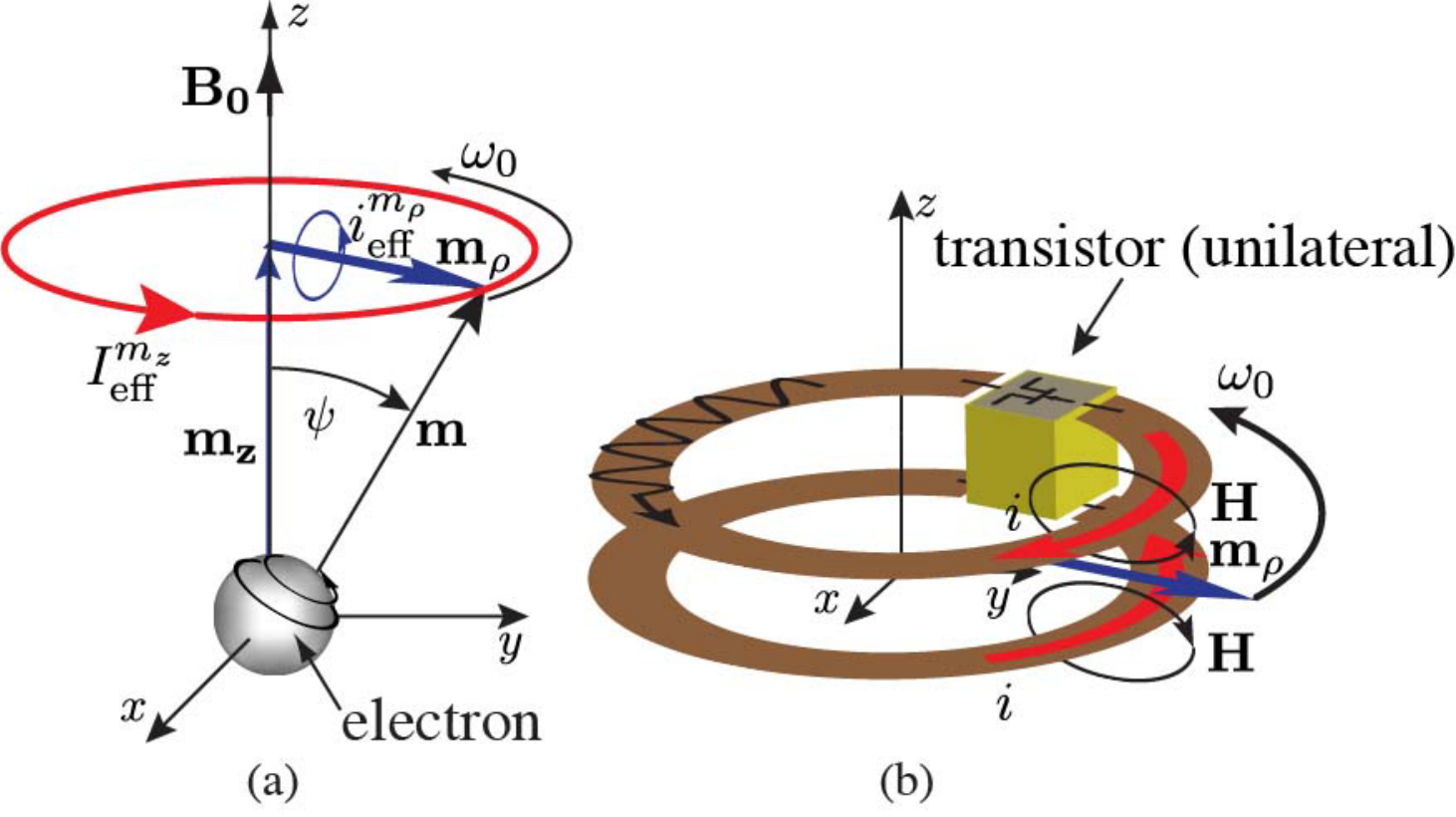}
\vspace{-3mm}
\caption{``Physicomimetic'' construction of the Unidirectional Loop Metamaterial (ULM) ``meta-molecule'' or particle. (a)~Magnetic dipole precession, arising from electron spinning in a ferrite material about the axis (here $z$) of an externally applied static magnetic bias field, $\ve{B}_0$, with effective unidirectional current loops $I_\text{eff}^{m_z}$ and $i_\text{eff}^{m_\rho}$, and transverse radial rotating magnetic dipole moment $\ve{m}_\rho$ associated with $i_\text{eff}^{m_\rho}$. (b)~ULM particle~\cite{Kodera_APL_2011}, typically (but not exclusively~\cite{Sounas_Alu_2014}) consisting of a pair of broadside-coupled transistor-loaded rings supporting antisymmetric current and unidirectional current wave (shown here with exaggeratedly small wavelength for the sake of visibility), with resulting radial rotating magnetic dipole moment emulating that in~(a).} \label{fig:physico_mimetics}\vspace{-8mm}
\end{figure}

Microwave magnetism in a ferrite is based on the \emph{precession of the magnetic dipole moments arising from unpaired electron spins about the axis of an externally applied static magnetic bias field,} $\ve{B}_0$, as illustrated in \figref{fig:physico_mimetics}(a), where $\ve{B}_0\|\uve{z}$. This is a quantum-mechanical phenomenon, that is described by the \emph{Landau-Lifshitz-Gilbert equation}~\cite{Landau_Lifshitz_1935,Lax_1962,Gurevich_1996}
\begin{equation}\label{eq:LLG}
\frac{d\bf{m}}{dt}
=-\gamma\ve{m}\times\ve{B}_0
+\frac{\alpha}{M_\text{s}}\ve{m}\times\frac{d\ve{m}}{dt},
\end{equation}
where $\ve{m}$ denotes the magnetic dipole moment, $\gamma$ the gyromagnetic ratio, $M_\text{s}$ the saturation magnetization, and $\alpha$ the Gilbert damping term. Equation~\eqref{eq:LLG} states that the \emph{time-variation rate of $\ve{m}$} due to a \emph{transverse}\footnote{The longitudinal ($z$) component does not contribute to precession, and hence to magnetism. Indeed, since $\ve{B_0}\|\uve{z}$, the $z$-component of $\ve{m}$ produced by $\ve{H}^\text{RF}_z$ would lead to $\ve{m}^\text{RF}_z\times(\ve{B}_0+\mu_0H^\text{RF}\uve{z})=[m^\text{RF}_z(B_0+\mu_0H^\text{RF})](\uve{z}\times\uve{z})=0$, the only torque being produced by the transverse component ($\ve{H}^\text{RF}_t$, $\uve{t}\in xy$-plane), $\ve{m}^\text{RF}_t\times(\ve{B}_0+\mu_0H^\text{RF}\uve{t})=(m^\text{RF}_tB_0)(\uve{t}\times\uve{z})\neq0$. In the rest of the text, we shall drop the superscript ``RF,'' without risk of ambiguity since $\ve{m}_t$, is exclusively produced by the RF signal.\label{fn:HRF_z}} RF magnetic field signal, $\ve{H}^\text{RF}_t$ ($\|\uve{t},\uve{t}\perp\uve{z}$), is equal to the sum of the \emph{torque} exerted by $\ve{B}_0$ on $\ve{m}$ (directed along $+\uvs{\phi}$, $\uvs{\phi}$: azimuth angle), and a \emph{damping term} (directed along $-\uvs{\theta}$, $\uvs{\theta}$: elevation angle) that reduces the precession angle, $\psi$, to zero along a circular-spherical trajectory (conserved $|\ve{m}|$) when the RF signal is suppressed (relaxation).

Classically, magnetic dipole moments can be associated with \emph{current loop sources}, according to Amp\`{e}re law. Decomposing a ferrite magnetic moment, $\ve{m}$, into its longitudinal component, $\ve{m}_z$, and transverse component, $\ve{m}_\rho$, as shown in \figref{fig:physico_mimetics}(a), one may thus invoke the effective current loops $I_\text{eff}^{m_z}$ and $i_\text{eff}^{m_\rho}$ as \emph{source models} for the corresponding moments. Among these currents, only $i_\text{eff}^{m_\rho}$ matters in terms of magnetism, since $I_\text{eff}^{m_z}$, as the source associated with $\ve{H}^\text{RF}_z$, does not induce any precession (Footnote~\ref{fn:HRF_z}). $i_\text{eff}^{m_\rho}$ is thus the current one has to mimic to devise an ``artificial ferrite.'' This current, as seen \figref{fig:physico_mimetics}, has the form of a loop tangentially rotating on an imaginary cylinder of axis~$z$.

Given its complexity, the current $i_\text{eff}^{m_\rho}$ may a priori seem impossible to emulate. However, what fundamentally matters for magnetism is not this current itself, but the moment $\ve{m}_\rho$, from which magnetization will arise at the macroscopic level (Sec.~\ref{sec:macro_descr}). This moment may be fortunately also produced by a pair of antisymmetric $\phi$-oriented currents, rotating on the same cylinder, which can be produced by a \emph{pair of conducting rings} operating in their odd mode~\cite{Sounas_TAP_01_2013}, as shown in \figref{fig:physico_mimetics}(b). If this ring-pair structure is loaded by a transistor~\cite{Kodera_APL_2011}, as depicted in the figure, or includes another \emph{unidirectionality mechanism} such as the injection of an azimuthal modulation~\cite{Sounas_Alu_2014}, $\ve{m}_\rho$ will \emph{unidirectionally} rotate about~$z$ when excited by an RF signal, and hence mimic the magnetic behavior of the electron in \figref{fig:physico_mimetics}(b). The structure in \figref{fig:physico_mimetics} constitutes thus the \emph{unit-cell particle} of the ULM at the \emph{microscopic level}.

One may argue there there is a fundamental difference between the physical system in \figref{fig:physico_mimetics}(a) and its presumed artificial emulation in \figref{fig:physico_mimetics}(b): the ferrite material also supports the longitudinal moment $\ve{m}_z$ whereas the ULM does not include anything alike. However, as we have just seen above, particularly in Footnote~\ref{fn:HRF_z}, $\ve{m}_z$ does not contribute to the magnetic response. It is therefore inessential and does thus not need to be emulated. So, the particle in \figref{fig:physico_mimetics}(b), with its moment $\ve{m}_\rho$ is all that is needed for \emph{artificial magnetism}!

Does this mean that the ULM particle includes no counterpart to the static alignment of the dipoles due to $\ve{B}_0$ (and producing $\ve{m}_z$) in the ferrite medium? In fact, there \emph{is} a counterpart, although $\ve{m}_z=0$ in the ULM. The fundamental outcome of the static alignment of the dipoles along $z$ in the ferrite is the alignment of the relevant magnetic dipoles $\ve{m}_\rho$ in the plane perpendicular to $\uve{z}$ (or within the $xy$-plane) across the medium, for otherwise the $\ve{m}_\rho$'s of the different domains~\cite{Lax_1962,Gurevich_1996} would macroscopically cancel out. Such an orientation of the $\ve{m}_\rho$'s perpendicularly to $\uve{z}$ is essential to emulate magnetism. How is this provided in the ULM? Simply by \emph{fixing the rings in a mechanical support}, such as a substrate, as will be seen later. So, the counterpart of the ferrite static alignment of dipoles is simply mechanical orientation in the ULM.

\subsection{Macroscopic Description}\label{sec:macro_descr}

\begin{figure}[htp]
\centering
\includegraphics[width=0.35\textwidth]{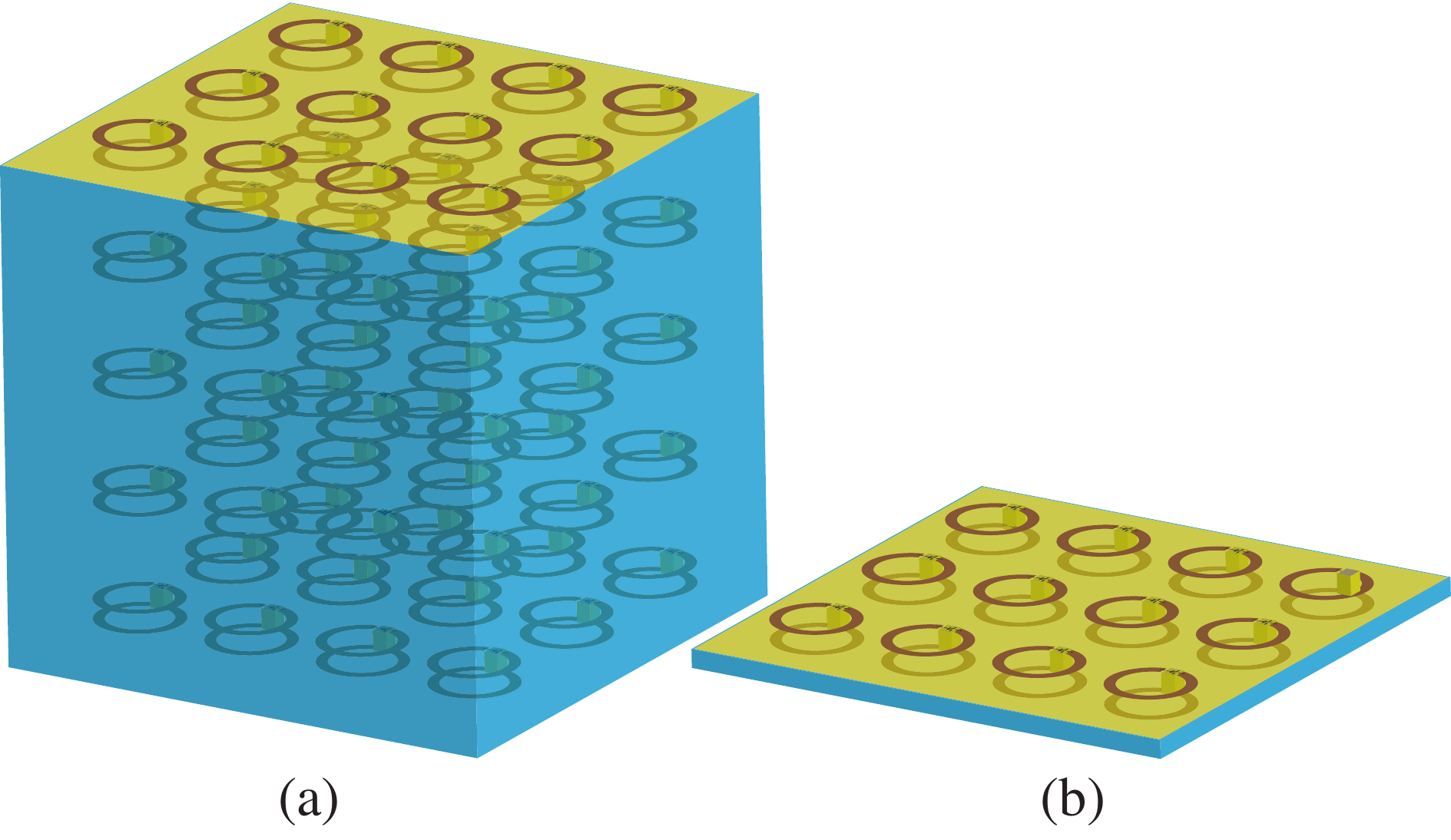}
\vspace{-3mm}
\caption{ULM structures obtained by periodically repeating the unit cell with the particle in Fig.~\ref{fig:physico_mimetics}. (a)~Metamaterial (3D), described by the Polder volume permeability~\eqref{eq:mu_tensor}. (b)~Metasurface (2D metamaterial), described by a surface permeability~\cite{Karim_2017_metasurface}.} \label{fig:metamaterial}
\end{figure}

Since it mimics the \emph{relevant magnetic} operation of a ferrite at the microscopic level, the unit-cell particle in \figref{fig:physico_mimetics}(b) must lead to the same response as bulk ferrite at the macroscopic level when repeated according to a subwavelength 3D lattice structure so as to form a metamaterial as shown in \figref{fig:metamaterial}(a).

The ULM in \figref{fig:metamaterial}(a), just as a ferrite\footnote{The difference is essentially \emph{quantitative}: while in the ferrite $p/\lambda<10^{-6}$ ($p$: molecular lattice constant), in the ULM $p/\lambda\approx 1/10-1/5$ ($p$: metamaterial lattice constant or period), but homogeneization works in both cases.}, forms a 3D array of magnetic dipole moments, $\ve{m}_i$, whose average over a subwavelength volume $V$,
\begin{equation}
\ve{M}
=\frac{1}{V}\sum_{i=1}\ve{m}_i
=\left(\frac{1}{V}\sum_{i=1}m_{\rho,i}\right)\uvs{\rho}
=M_\rho\uvs{\rho}
\end{equation}
corresponds to the density of magnetic dipole moments, or \emph{magnetization}, as the fundamental macroscopic quantity describing the metamaterial\footnote{Whereas in a ferrite, we have $\ve{M}=\ve{M}_\text{s}+\ve{M}^\text{RF}=(M_\text{s}+M_z^\text{RF})\uve{z}+\ve{M}_t^\text{RF}\approx M_\text{s}\uve{z}+M_\rho^\text{RF}\uvs{\rho}$, where $M_\text{s}$ is the saturation magnetization of material, in the ULM $M_s=0$. We shall subsequently drop the superscript ``RF'' also in $M_\rho$.}.

\vspace{-3mm}

From this point, one may follow the same procedure as in ferrites~\cite{Polder_1949, Pozar_ME_2011} to obtain the Polder ULM permeability tensor
\begin{subequations}\label{eq:mu_overall}
\begin{equation}\label{eq:mu_tensor}
\overline{\overline{\bf{\mu}}}=\begin{bmatrix}
\mu & j\kappa & 0 \\
-j\kappa & \mu & 0 \\
0 & 0 & \mu_0,
\end{bmatrix},
\end{equation}
\begin{equation}\label{eq:mu_kappa}
\text{with}\;\mu=\mu_0\left(1+\frac{\omega_0\omega_\text{m}}{\omega_0^2-\omega^2}\right)
\;
\text{and}\;\kappa=\mu_0\frac{\omega\omega_\text{m}}{\omega_0^2-\omega^2},
\end{equation}
\end{subequations}
where $\omega_0$ and $\omega_\text{m}$ are the ULM \emph{resonance frequency} (or Larmor frequency) and \emph{effective saturation magnetization frequency}, respectively\footnote{In a ferrite, $\omega_0=\gamma B_0$ and $\omega_\text{m}=\gamma\mu_0M_\text{s}$.}, that will be derived in the next section. As in ferrites, the effect of loss can be accounted for by the substitution $\omega_0\leftarrow\omega_0+j\alpha\omega$, where $\alpha$ a damping factor in~\eqref{eq:LLG}~\cite{Pozar_ME_2011}.

So, a ULM may really be seen as a an artificial ferrite material producing \emph{magnet-less} \emph{artificial magnetism}. However, its \emph{nonreciprocity} is achieved from \emph{breaking time-reversal (TR) symmetry} by a \emph{TR-odd current bias}, originating in the transistor (DC) biasing, instead of a TR-odd external magnetic field~\cite{Caloz_AWPL_NR_I_2018}.

ULMs have been implemented only in a 2D format so far. The corresponding structure is shown in \figref{fig:metamaterial}(b), and may be referred to as a \emph{Unilateral Loop Metasurface (ULMS)}. Section~\ref{sec:Faraday_rot} will present ULMS \emph{Faraday rotation} and Sec.~\ref{sec:MS_appl} will discuss related applications.

\section{ULM Particle and Design}\label{sec:impl_des}
\vspace{-1mm}

\begin{figure}[htp]
\centering
\includegraphics[width=0.27\textwidth]{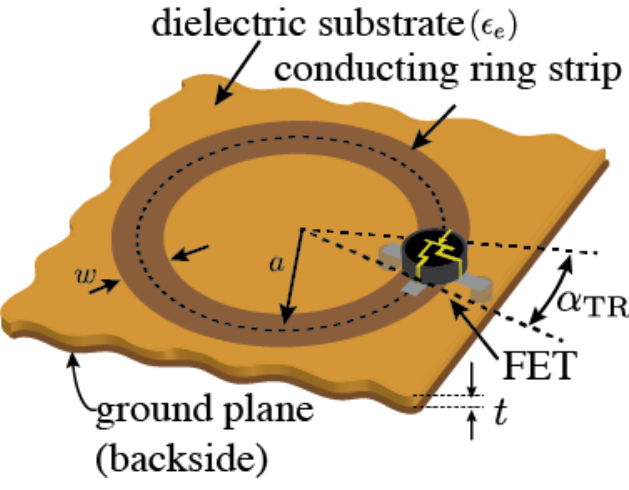}
\vspace{-3mm}
\caption{ULM particle microstrip implementation in the form of a transistor-loaded single ring on a PEC plane, supporting the odd effective current distribution and unidirectional current wave as in \figref{fig:physico_mimetics}(b). The transistor biasing circuit is not shown here.}
\label{fig:design_fig}
\end{figure}

ULMs may be implemented in different manners. Figure~\ref{fig:design_fig} shows a ULM particle implemented in the form of a microstrip transistor-loaded single ring placed on PEC plane. Assuming a distance much smaller than the wavelength between the ring and the PEC plane, the structure is equivalent, by the image principle, to the antisymmetric double-ring structure in \figref{fig:physico_mimetics}(b)~\cite{Sounas_TAP_01_2013}.

\vspace{-6.7mm}

The transistor-loaded ULM particle in \figref{fig:design_fig}, or \figref{fig:physico_mimetics}(b), is essentially a \emph{ring resonator}, whose total electrical size is given by~\cite{Kodera_TMTT_03_2013}
\begin{equation}\label{eq:ring_phase}
\beta_\text{ms}a(2\pi-\alpha_\text{TR})+\varphi_\text{TR}=2\pi,
\;
\beta_\text{ms}=k_0\sqrt{\epsilon_\text{e}}=\dfrac{\omega}{c}\sqrt{\epsilon_\text{e}},
\end{equation}
where $\beta_\text{ms}$ is the microstrip line wavenumber ($\epsilon_\text{e}$: effective relative permittivity), $a$ is the average radius of the ring, $\alpha_\text{TR}$ is the geometrical angle subtending the transistor chip, and $\varphi_\text{TR}$ is the phase shift across it. Solving Eq.~\eqref{eq:ring_phase} for $\omega$ provides the resonance frequency of the resonator, and hence the \emph{ULM resonance frequency},
\vspace{-2mm}
\begin{equation}
\omega_0=\left| \frac{(2\pi-\varphi_\text{TR})c}{a\sqrt{\epsilon_\text{e}}(2\pi-\alpha_\text{TR})}\right|,
\label{eq:res_omega}
\end{equation}
in~\eqref{eq:mu_overall}. The parameter $\omega_\text{m}$ in the same relations follows from the mechanical orientation of the moments, as explained in Sec.~\ref{sec:macro_descr}: although we do not have here a saturation magnetization $M_\text{s}$ leading to the frequency parameter $\omega_\text{m}=\gamma\mu_0M_\text{m}$ in the ferrite, we have have an equivalent phenomenological parameter $\omega_\text{m}$ associated with the orientation of the rings, which may be found by extraction, as will be seen in Sec.~\ref{sec:Faraday_rot}.

\begin{figure}[htp]
\centering
\includegraphics[width=0.4\textwidth]{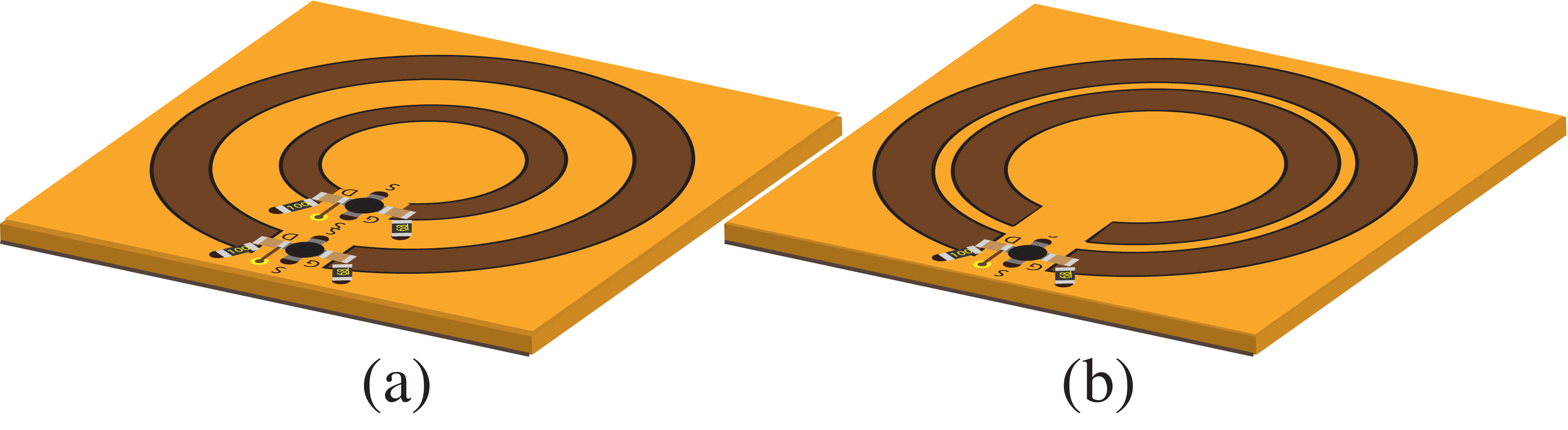}
\vspace{-3mm}
\caption{Unique magnetic properties of the ULM attainable by using multiple rings of resonance frequencies $\omega_{0n}$, here two rings with resonance frequencies $\omega_{01}$ and $\omega_{02}$. (a)~Multiband operation using independent rings with separate resonance frequencies. (b)~Enhanced bandwidth using coupled-resonant rings with overlapping resonance frequencies.} \label{fig:enhance}
\end{figure}

Note that ULMs may be designed for \emph{multi-band} operation and \emph{enhanced-bandwidth} operation. The former, in contrast to ferrites that are restricted to a single ferromagnetic resonance $\omega_0=\gamma B_0$\footnote{This restriction can be somewhat overcome in a structured ferromagnetic structure, such as a ferromagnetic nanowire membrane supporting a remanent bistable population of up and down magnetic dipole moments with corresponding resonances $\omega_{0}^{\uparrow}=\gamma\mu_0H_0^{\uparrow}$ and $\omega_{0}^{\downarrow}=\gamma\mu_0H_0^{\downarrow}$ \cite{Carignan_2009, Boucher_2009}.}, can in principle accommodate multiple resonances by simply incorporating rings of different sizes, as illustrated in Fig.~\ref{fig:enhance}. The latter, in contrast to ferrite whose bandwidth is inversely proportional to loss due to causality, can be achieved by leveraging overlapping coupled resonances~\cite{Matthaei_gyrator_1959}.
\vspace{-3mm}

The single-ring-PEC ULM implementation of \figref{fig:design_fig} is ideal for microstrip components~\cite{Kodera_TMTT_03_2013} and reflective metasurfaces~\cite{Kodera_APL_2011}. However, for 3D ULM [\figref{fig:metamaterial}(a)] structures and transmissive metasurfaces (Sec.~\ref{sec:Faraday_rot}), a transparent version of that ULM is required. This could theoretically be realized with a pair of rings, as in \figref{fig:physico_mimetics}(b), but may be more conveniently implemented in the form of circular slots in a Coplanar Wave-Guide (CPW) type technology, as reported in~\cite{Kodera_E_type_2012}.

\section{Faraday Rotation}\label{sec:Faraday_rot}
\vspace{-1mm}

Faraday rotation is one of the most fundamental and useful properties of magnetic materials. Given their artificial ferrite nature (Sec.~\ref{sec:impl_des}), ULMs can readily support this effect. The \emph{Faraday angle} is given by~\cite{Lax_1962,Pozar_ME_2011}
\begin{equation}\label{eq:rotation_ang}
\theta_\ve{F}(z)=-\left( \frac{\beta_+-\beta_-}{2}z \right),
\;\text{with}\;
\beta_{\pm}=\omega\sqrt{\epsilon (\mu \pm \kappa)},
\end{equation}
where $\mu$ and $\kappa$ are the Polder tensor components in \eqref{eq:mu_kappa}, with the resonance ($\omega_0$) given by \eqref{eq:res_omega} and the saturation magnetization frequency ($\omega_\text{m}$) discussed in Sec.~\ref{sec:impl_des}. Interesting, the ULM allows the option to \emph{reverse the direction of Faraday rotation by simple voltage control} (instead of magnet mechanical flipping in a conventional ferrite) using an antiparallel transistor pair load, as demonstrated in~\cite{Kodera_AWPL_12_2012}.

\begin{figure}[htp]
\centering
\includegraphics[width=0.43\textwidth]{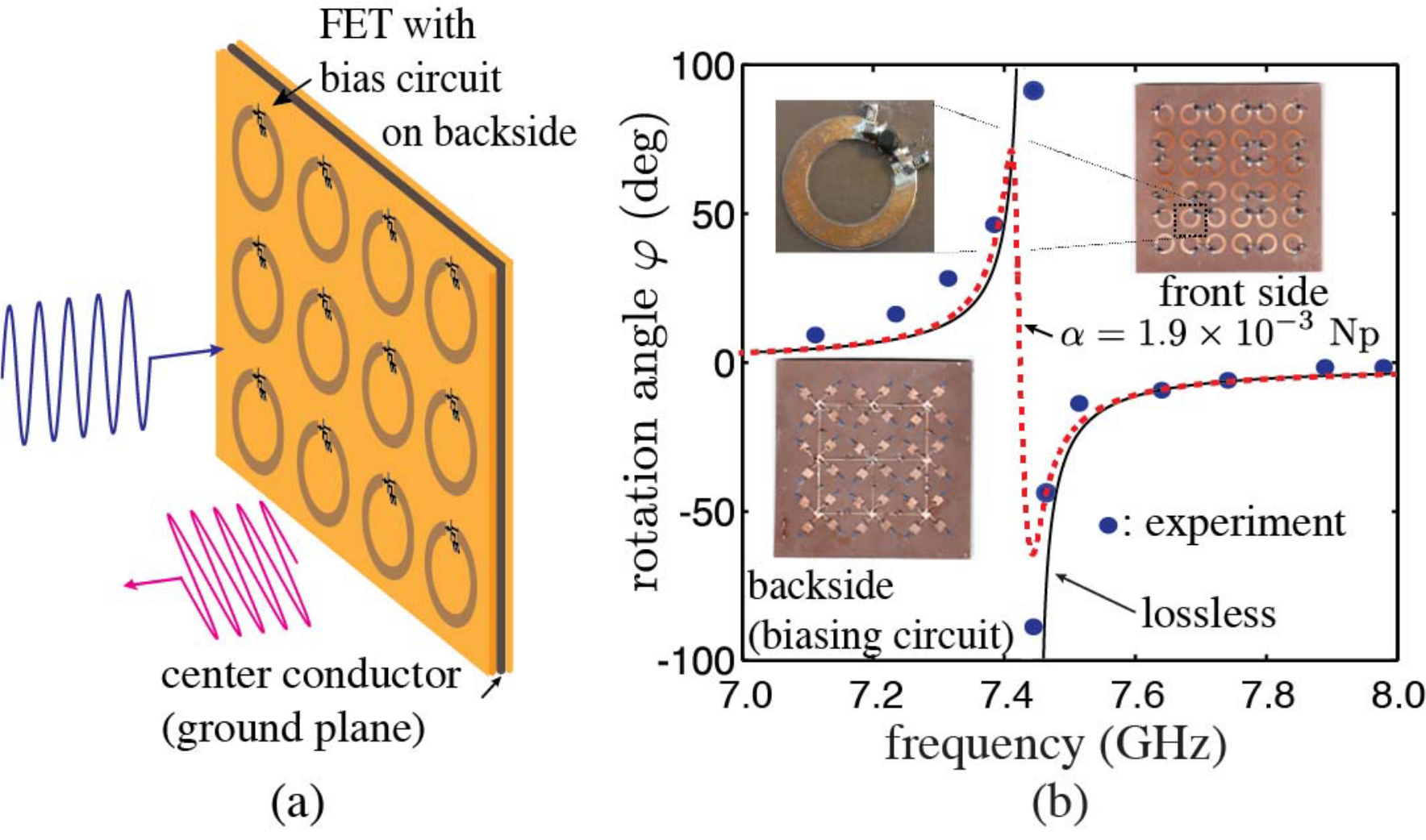}
\vspace{-3mm}
\caption{Reflective Faraday rotating ULMS based on the particle in \figref{fig:design_fig}~\cite{Kodera_APL_2011}. (a)~Perspective representation of the metasurface with rotated plane of polarization. (b)~Theoretical [Eq.~\eqref{eq:rotation_ang}] and experimental polarization rotation angle versus frequency. Here, $\epsilon_\text{r}=2.6$, $\omega_0/2\pi=7.42$~GHz ($\ve{B}_0^\text{equiv.}=0.265$~T), $\omega_\text{m}/2\pi=28$~MHz ($\mu_0\ve{M}_\text{s}^\text{equiv.}=1$~mT) and $\alpha=1.9\times10^{-3}$~Np ($\Delta H=\alpha \omega_0/\gamma \mu_0=0.4$ mT.)} \label{fig:rotator}
\end{figure}

Figure~\ref{fig:rotator} shows a reflective Faraday ULM metasurface (ULMS) structure, based on the particle in \figref{fig:design_fig}, and response, initially reported in~\cite{Kodera_APL_2011}. The results confirm that the ULM works exactly as a ferrite, whose equivalent parameters are given in the caption.

\vspace{-2mm}

ULM Faraday rotation has also been reported in transmission, using the circular-slot ULM structure mentioned at the end of Sec.~\ref{sec:impl_des}. Using slots, and hence equivalent magnetic currents, instead of rings supporting electric currents, that structure really operates as an artificial \emph{magneto-optic} material, with a \emph{permittivity tensor} replacing the magnetic tensor in~\eqref{eq:mu_tensor}. A similar Faraday rotation effect may also be achieved using arrays of twisted dipoles loaded by transistors~\cite{Joannopoulos_2014}.

\section{Applications}
\vspace{-1mm}

\subsection{Metasurface Isolators}\label{sec:MS_appl}
\vspace{-2mm}

\begin{figure}[htp]
\centering
\includegraphics[width=0.34\textwidth]{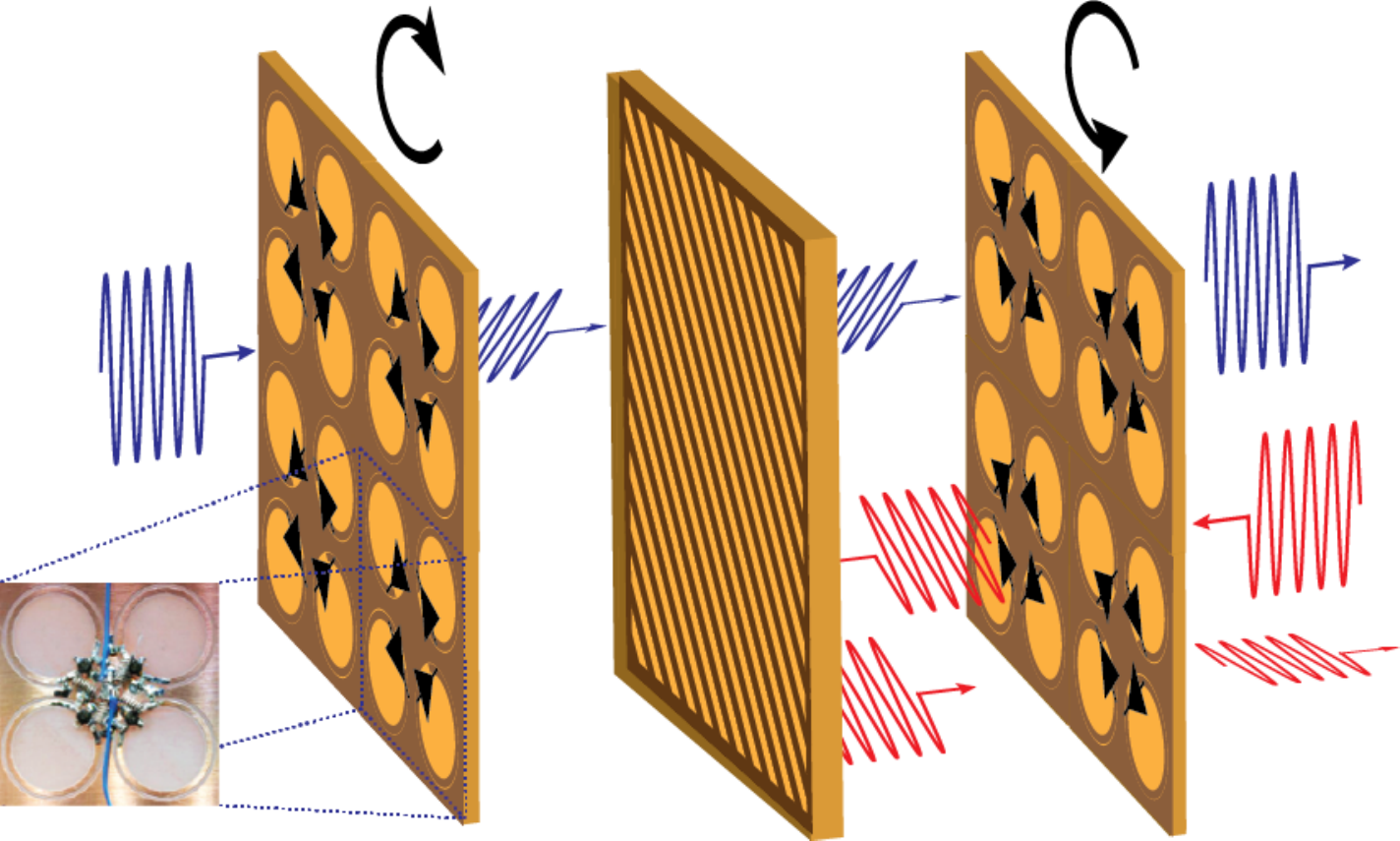}
\vspace{-2mm}
\caption{Isolator using two transmissive Faraday rotation ULMSs~\cite{Kodera_E_type_2012} and a 45$^\circ$ polarizer. The bottom-left inset shows the transmissive ``magneto-optic'' slot-ULM demonstrated in~\cite{Kodera_E_type_2012}, that may be used for this application.} \label{fig:isolator_by_rotator}
\end{figure}

The transmissive ULMS in~\cite{Kodera_E_type_2012} can be straightforwardly applied to build a Faraday isolator~\cite{Hogan_Faraday_isolator_1952,Lax_1962,Optical_Faraday_iso_1964,Rosenberg_opt_iso_1964}, as shown in \figref{fig:isolator_by_rotator}. As the wave propagates from the left to the right, its polarization is rotated 45$^\circ$ by the left ULMS in the rotation direction imposed by the transistors (here, clock-wise). It thus reaches the polarizer with its electric field perpendicular to the conducting strips and therefore unimpededly crosses it. It is finally rotated back to its initial (vertical) direction by the right ULMS, whose rotation direction is opposite to the left one (here, counter-clockwise). In the opposite direction, the right ULMS rotates the wave polarization in such a manner that its electric field is parallel to the conducting strips of the polarizer, so that the wave is completely reflected. It is then rotated again by the right polarizer and gets back to the right input orthogonal to the original wave\footnote{Lossy polarizers can be added, if necessary (The final wave could be reflected back), for dissipative (rather than reflective) isolation~\cite{Parsa_TAP_03_2011}.}.
%

Faraday rotation is not the only approach to realize \emph{spatial isolation}, as in \figref{fig:isolator_by_rotator}. Such isolation may be simply achieved, without any gyrotropy but still magnetlessly, with a metasurface consisting of back-to-back antenna arrays interconnected by transistors~\cite{Taravati_TAP_07_2017}; this nonreciprocal metasurface may exhibit an ultra wideband response and provide transmission gain.

\vspace{-1.5mm}
\subsection{Nonreciprocal Antenna Systems}
\begin{figure}[htp]
\centering
\includegraphics[width=0.45\textwidth]{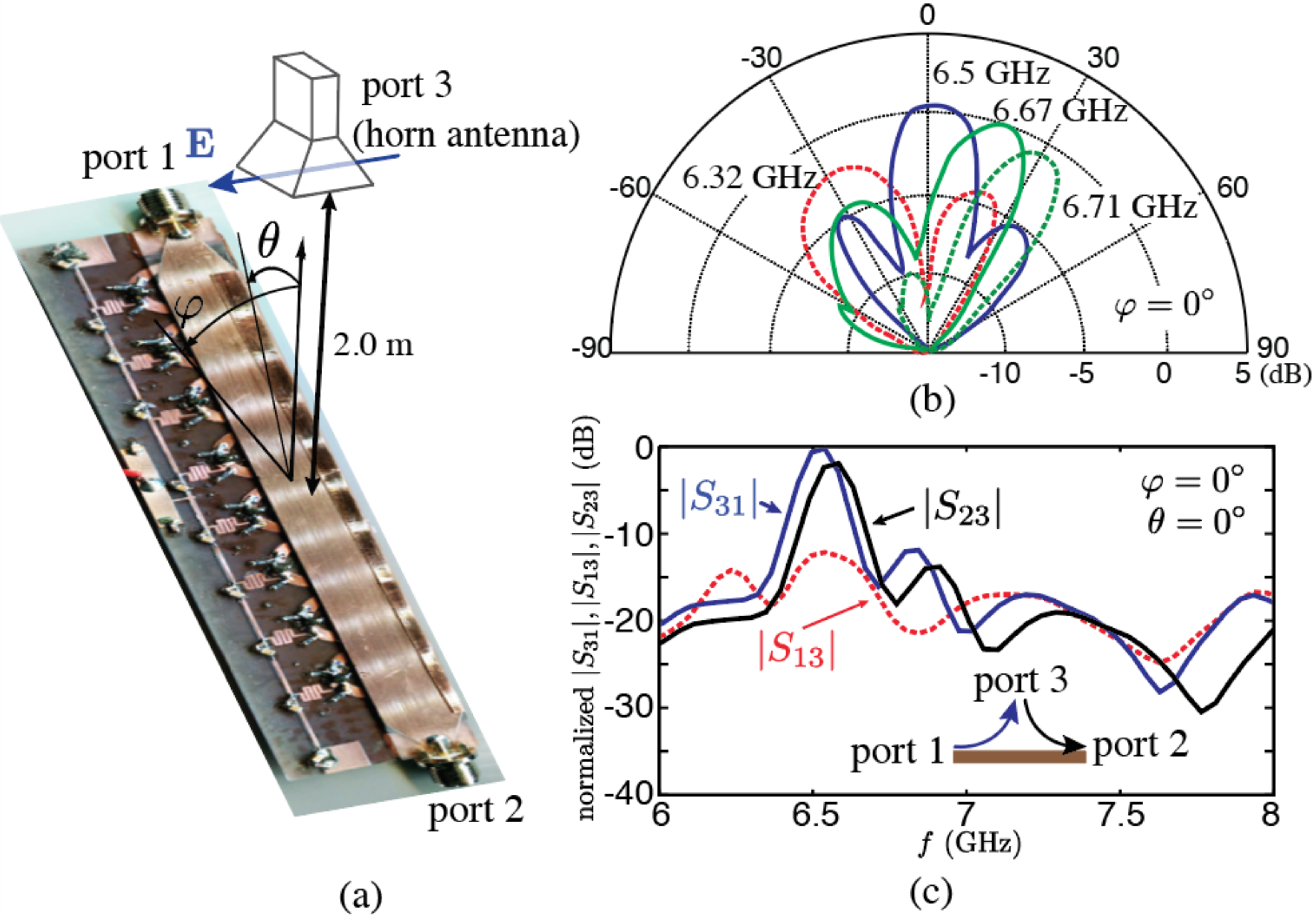}
\vspace{-4mm}
\caption{ULM CRLH leaky-wave isolated-antenna or antenna-duplexer system~\cite{Kodera_AWPL_01_2012}. (a)~Prototype with port definitions. (b)~Measured radiation patterns. (c)~Measured scattering parameters.} \vspace{-5mm}
\label{fig:CRLH_ant}
\end{figure}

\vspace{-1.5mm}
The ULM structure may be used in various nonreciprocal radiating (antenna, reflector and metasurface) applications. Figure~\ref{fig:CRLH_ant} shows a nonreciprocal antenna system and its response~\cite{Kodera_AWPL_01_2012}. The structure [\figref{fig:CRLH_ant}(a)] is a ULM magnet-less version of the nonreciprocal ferrite-loaded Composite Right/Left-Handed (CRLH) open-waveguide leaky-wave antenna introduced in~\cite{ferrite_CRLH_2009,Kodera_TAP_10_2010}, with the ferrite material replaced by a 1D ULM structure. This structure may be used as a \emph{nonreciprocal full-space scanning antenna} [Figs.~\ref{fig:CRLH_ant}(b) and (c)], whose unidirectionality provides protection against interfering signals, or as an \emph{antenna diplexer system}, where nonreciprocity effectively plays the role of a circulator with highly isolated uplink ($3\rightarrow 1$) and downlink ($2\rightarrow 3$) paths.\vspace{-1mm}
\vspace{-2mm}

\subsection{Isolators and Circulators}
\begin{figure}[htp]
\centering
\includegraphics[width=0.39\textwidth]{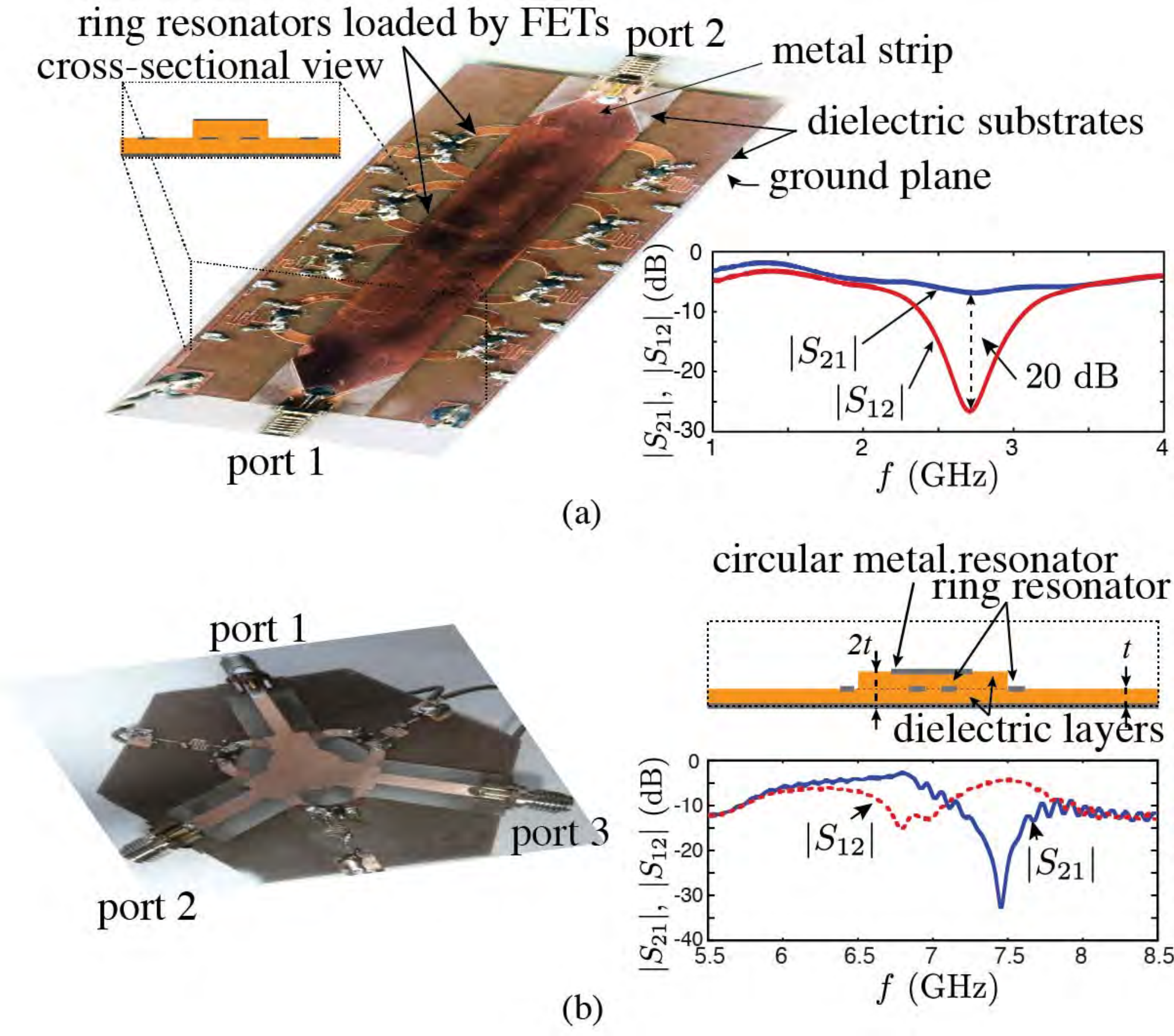}
\vspace{-4mm}
\caption{ULM components~\cite{Kodera_TMTT_03_2013}. (a)~Isolator\cite{Frustrated_2012}. (b)~Circulator~\cite{Kodera_TMTT_03_2013}.} \label{fig:application}
\vspace{-4mm}
\end{figure}

\vspace{-2mm}
ULM technology also enables various kinds of nonreciprocal components. Figure~\ref{fig:application}(a) shows a ULM microstrip \emph{isolator}~\cite{Kodera_TMTT_03_2013}. The ULM structure below the microstrip line is composed of two rows of transistor-loaded rings with opposite biasing, and hence opposite allowed wave rotation directions. As the wave from the microstrip line reaches a ring pair, its mode is coupled into a stripline mode with strip pair constituted by the longitudinal sections of the overlapping rings, and usual antisymmetric currents. In the propagation direction where these currents are co-directional, with allowed rotation direction of the ULM, the stripline mode is allowed to propagate, whereas in the opposite propagation direction, it is inhibited and dissipates in matching resistors on the rings. Figure~\ref{fig:application}(b) shows a ULM microstrip \emph{circulator}~\cite{Kodera_TMTT_03_2013}, which is based on mode-split counter-rotating modes as all circulators.
\vspace{-6mm}

\section{Conclusion}
\vspace{-2mm}

We have presented an overview of ULM structures and applications. The ULM physics has been described in great details, revealing that the ULM really represents an artificial ferrite medium. It is in fact the only existing medium of the kind. It has been pointed out that the ULM may offer unique extra benefits compared to ferrites, such a multiband operation, ultra broadband and electronic Faraday rotation direction switching.

\clearpage

\bibliographystyle{IEEEtran}
\bibliography{main}
\end{document}